\documentclass[aps,preprint,nofootinbib,superscriptaddress]{revtex4}%
\usepackage{amssymb}
\usepackage{amsmath}
\usepackage{epsfig}
\usepackage{amsfonts}
\usepackage{graphicx}
\usepackage{hyperref}%
\providecommand{\U}[1]{\protect\rule{.1in}{.1in}}
\begin{document}
\title{Yang-Mills Instanton Sheaves}
\author{Sheng-Hong Lai}
\email{xgcj944137@gmail.com}
\affiliation{Department of Electrophysics, National Chiao-Tung University, Hsinchu, Taiwan, R.O.C.}
\author{Jen-Chi Lee}
\email{jcclee@cc.nctu.edu.tw}
\affiliation{Department of Electrophysics, National Chiao-Tung University, Hsinchu, Taiwan, R.O.C.}
\author{I-Hsun Tsai}
\email{ihtsai@math.ntu.edu.tw}
\affiliation{Department of Mathematics, National Taiwan University, Taipei, Taiwan, R.O.C. }
\date{\today}

\begin{abstract}
The $SL(2,C)$ Yang-Mills instanton solutions constructed recently by the
biquaternion method were shown to satisfy the complex version of the ADHM
equations and the monad construction. Moreover, we discover that, in addition
to the holomorphic vector bundles on $CP^{3}$ similar to the case of $SU(2)$
ADHM construction, the $SL(2,C)$ instanton solutions can be used to explicitly
construct instanton sheaves on $CP^{3}$. Presumably, the existence of these
instanton sheaves is related to the singularities of the $SL(2,C)$ instantons
on $S^{4}$ which do not exist for $SU(2)$ instantons.

\end{abstract}
\maketitle
\tableofcontents

%

%TCIMACRO{\TeXButton{equation number}{\setcounter{equation}{0}
%\renewcommand{\theequation}{\arabic{section}.\arabic{equation}}}}%
%BeginExpansion
\setcounter{equation}{0}
\renewcommand{\theequation}{\arabic{section}.\arabic{equation}}%
%EndExpansion

\section{Introduction}

Since the discovery of classical exact solutions of Euclidean $SU(2)$
(anti)self-dual Yang-Mills (SDYM) equation in 1970s, there has been many
important and interesting research activities on YM instantons both in quantum
field theory and algebraic geometry. On the physics side, applications of
quantum instanton tunnelling in nonperturbative quantum field theory has
resolved the QCD $U(1)_{A}$ problem \cite{U(1)} and, on the other hand,
created the strong $CP$ problem with associated QCD $\theta$-vacua \cite{the}
structure. Mathematically, one important application of instantons in
differential topology was the classification of four-manifolds \cite{5}.

The first BPST $1$-instanton solution \cite{BPST} was found in 1975. Soon
later the CFTW $k$-instanton solutions \cite{CFTW} with $5k$ moduli parameters
were constructed, and then the number of moduli parameters of the
$k$-instanton solutions was extended to $5k+4$ ($5$,$13$ for $k=1$,$2$)
\cite{JR} based on the $4D$ conformal symmetry group. Finally, the complete
solutions with $8k-3$ moduli parameters for each $k$-th homotopy class were
worked out in 1978 by mathematicians ADHM \cite{ADHM} using method in
algebraic geometry. By using the monad construction combining with the
Penrose-Ward transform, ADHM constructed the most general instanton solutions
by establishing an one to one correspondence between anti-self-dual
$SU(2)$-connections on $S^{4}$ and global holomorphic vector bundles of rank
two on $CP^{3}$. The explicit closed forms of the complete $SU(2)$ instanton
solutions for $k\leq3$ had been worked out in \cite{CSW}.

In addition to $SU(2)$, the ADHM construction has been generalized to the
cases of $SU(N)$ and many other SDYM theories with compact Lie groups
\cite{CSW,JR2}. In a recent paper \cite{Ann}, the present authors generalized
the quaternion calculation of $SU(2)$ ADHM construction to the biquaternion
calculation with biconjugation operation, and constructed a class of
non-compact $SL(2,C)$ YM instanton solutions with $16k-6$ parameters for each
$k$-th homotopy class. The number of parameters is consistent with the
conjecture made by Frenkel and Jardim in \cite{math2} and was proved recently
in \cite{math3} from the mathematical point of view. These new $SL(2,C)$
instanton solutions contain previous $SL(2,C)$ $(M,N)$ instanton solutions as
a subset constructed in 1984 \cite{Lee}.

One important motivation to study $SL(2,C)$ instanton solutions has been to
understand, in addition to the holomorphic vector bundles on $CP^{3}$ in the
ADHM construction which has been well studied in the $SU(2)$ instantons, the
instanton sheaf structure on the projective space. Indeed, in constrast to the
well known $SU(2)$ regular instanton solutions without singularities on
$S^{4}$ spacetime, it was discovered that \cite{Ann} there were singularities
for $SL(2,C)$ instanton solutions on $S^{4}$ which can not be gauged away. Now
recall that there is a fibration from $CP^{3}$ to $S^{4}$ with fibers being
$CP^{1}$. A bundle $E$ on $CP^{3}$ can descend down to a bundle over $S^{4}$
if and only if no fiber of the twistor fibration is a jumping line for $E$.
This is precisely the case for the $SU(2)$ ADHM construction.

For the case of $SL(2,C)$ instanton solutions, things get more interesting.
Since some twistor lines are jumping lines, one may expect the existence of
$SL(2,C)$ instanton sheaf structure on $CP^{3}$ after Penrose-Ward transform
in the ADHM construction$.$ In this paper, we will show that for the $SL(2,C)$
CFTW $k$-instanton solutions with $10k$ moduli parameters, although the
jumping lines exist on $S^{4}$, the vector bundle description on $CP^{3}$
remains valid as in the case of $SU(2)$ instantons. We then proceed to
calculate the case of more general known $SL(2,C)$ $2$-instanton solutions
with $26$ moduli parameters. We discover that, for some points on $CP^{3}$ and
some subset of the $26D$ moduli space of $2$-instanton solutions, the vector
bundle description of $SL(2,C)$ $2$-instanton on $CP^{3}$ breaks down, and one
is led to use a description in terms of torsion free sheaves for these
non-compact YM instantons or "instanton sheaves" on $CP^{3}$\cite{math2}.

This paper is organized as following. In section II, we review the
biquaternion construction of $SL(2,C)$ YM instantons with $16k-6$ moduli
parameters \cite{Ann}. In section III, we show that the $SL(2,C)$ YM instanton
solutions constructed in section II are solutions of the complex version of
the ADHM equations \cite{Donald}. Along with the calculation, we identify the
complex ADHM data $(B_{lm},I_{m,}J_{m})$ with $l,m=1,2.$ We then identify the
corresponding $\alpha$ and $\beta$ matrices in the monad construction of the
holomorphic vector bundles on $CP^{3}$. In section IV, we show that the
$SL(2,C)$ CFTW instanton solutions with $10k$ parameters on $S^{4}$ correspond
to the locally free sheaves or holomorphic vector bundles on $CP^{3}.$ We then
examine the locally free conditions \cite{math2} of the complete $SL(2,C)$
$2$-instanton solutions with $26$ moduli parameters, and discover the
existence of the $SL(2,C)$ $2$-instanton sheaves on $CP^{3}$ for some subset
of the $26D$ moduli space. Finally, a brief conclusion is given in section V.

\section{Biquaternions and $SL(2,C)$ Instantons}

In this section, we review biquaternion construction of $SL(2,C)$ YM instanton
solutions calculated in \cite{Ann}. We begin with the discussion of $SL(2,C)$
YM equation. There are two linearly independent choices of $SL(2,C)$ group
metric \cite{WY}
\begin{equation}
g^{a}=%
\begin{pmatrix}
I & 0\\
0 & -I
\end{pmatrix}
,g^{b}=%
\begin{pmatrix}
0 & I\\
I & 0
\end{pmatrix}
\end{equation}
where $I$ is the $3\times3$ unit matrix. In general, one can choose
\begin{equation}
g=\cos\theta g^{a}+\sin\theta g^{b}%
\end{equation}
where $\theta$ = real constant. Note that this metric is not positive definite
due to the non-compactness of $SL(2,C).$ On the other hand, it can be shown
that, as a differential manifold, $SL(2,C)$ is isomorphic to $S^{3}\times
R^{3},$ and one can easily calculate its third homotopy group \cite{Lee}%
\begin{equation}
\pi_{3}[SL(2,C)]=\pi_{3}[S^{3}\times R^{3}]=\pi_{3}(S^{3})\cdot\pi_{3}%
(R^{3})=Z\cdot I=Z\newline\newline%
\end{equation}
\newline where $I$ is the identity group, and $Z$ is the integer group.

Wu and Yang \cite{WY} have shown that a complex $SU(2)$ gauge field is related
to a real $SL(2,C)$ gauge field. Starting from $SU(2)$ complex gauge field
formalism, we can write down all the $SL(2,C)$ field equations. Introduce the
complex gauge field
\begin{equation}
G_{\mu}^{a}=A_{\mu}^{a}+iB_{\mu}^{a},
\end{equation}
the corresponding complex field strength is defined as ($g=1$)
\begin{equation}
F_{\mu\nu}^{a}\equiv H_{\mu\nu}^{a}+iM_{\mu\nu}^{a},a,b,c=1,2,3
\end{equation}
where
\begin{align}
H_{\mu\nu}^{a}  &  =\partial_{\mu}A_{\nu}^{a}-\partial_{\nu}A_{\mu}%
^{a}+\epsilon^{abc}(A_{\mu}^{b}A_{\nu}^{c}-B_{\mu}^{b}B_{\nu}^{c}),\nonumber\\
M_{\mu\nu}^{a}  &  =\partial_{\mu}B_{\nu}^{a}-\partial_{\nu}B_{\mu}%
^{a}+\epsilon^{abc}(A_{\mu}^{b}B_{\nu}^{c}-A_{\mu}^{b}B_{\nu}^{c}).
\end{align}
The $SL(2,C)$ YM equation can then be written as
\begin{align}
\partial_{\mu}H_{\mu\nu}^{a}+\epsilon^{abc}(A_{\mu}^{b}H_{\mu\nu}^{c}-B_{\mu
}^{b}M_{\mu\nu}^{c})  &  =0,\nonumber\\
\partial_{\mu}M_{\mu\nu}^{a}+\epsilon^{abc}(A_{\mu}^{b}M_{\mu\nu}^{c}-B_{\mu
}^{b}H_{\mu\nu}^{c})  &  =0,
\end{align}
and the $SL(2,C)$ SDYM equations are%
\begin{align}
H_{\mu\nu}^{a}  &  =\frac{1}{2}\epsilon_{\mu\nu\alpha\beta}H_{\alpha\beta
},\nonumber\\
M_{\mu\nu}^{a}  &  =\frac{1}{2}\epsilon_{\mu\nu\alpha\beta}M_{\alpha\beta}.
\label{self}%
\end{align}
YM equation for the choice $\theta=0$ can be derived from the following
Lagrangian
\begin{equation}
L=\frac{1}{4}(H_{\mu\nu}^{a}H_{\mu\nu}^{a}-M_{\mu\nu}^{a}M_{\mu\nu}^{a}).
\end{equation}

\bigskip We now proceed to review the construction of $SL(2,C)$ YM instantons
\cite{Lee,Ann}. We will use the convention $\mu=0,1,2,3$ and $\epsilon
_{0123}=1$ for $4D$ Euclidean space. In contrast to the quaternion in the
$Sp(1)$ ($=SU(2)$) ADHM construction, the authors of \cite{Ann} used
\textit{biquaternion} to construct $SL(2,C)$ YM instantons. A quaternion $x$
can be written as%
\begin{equation}
x=x_{\mu}e_{\mu}\text{, \ }x_{\mu}\in R\text{, \ }e_{0}=1,e_{1}=i,e_{2}%
=j,e_{3}=k \label{x}%
\end{equation}
where $e_{1},e_{2}$ and $e_{3}$ anticommute and obey%
\begin{align}
e_{i}\cdot e_{j}  &  =-e_{j}\cdot e_{i}=\epsilon_{ijk}e_{k};\text{
\ }i,j,k=1,2,3,\\
e_{1}^{2}  &  =-1,e_{2}^{2}=-1,e_{3}^{2}=-1.
\end{align}
The conjugate quarternion is defined to be%

\begin{equation}
x^{\dagger}=x_{0}e_{0}-x_{1}e_{1}-x_{2}e_{2}-x_{3}e_{3}%
\end{equation}
so that the norm square of a quarternion is%
\begin{equation}
|x|^{2}=x^{\dagger}x=x_{0}^{2}+x_{1}^{2}+x_{2}^{2}+x_{3}^{2}. \label{norm}%
\end{equation}
Occasionaly the unit quarternions can be expressed as Pauli matrices%
\begin{equation}
e_{0}\rightarrow%
\begin{pmatrix}
1 & 0\\
0 & 1
\end{pmatrix}
,e_{i}\rightarrow-i\sigma_{i}\ \text{; }i=1,2,3.
\end{equation}

A biquaternion (or complex-quaternion) $z$ can be written as%
\begin{equation}
z=z_{\mu}e_{\mu}\text{, \ }z_{\mu}\in C,
\end{equation}
which occasionally can be written as%
\begin{equation}
z=x+yi
\end{equation}
where $x$ and $y$ are quaternions and $i=\sqrt{-1},$ not to be confused with
$e_{1}$ in Eq.(\ref{x}). The biconjugation \cite{Ham} of $z$ is defined to be%
\begin{equation}
z^{\circledast}=z_{\mu}e_{\mu}^{\dagger}=z_{0}e_{0}-z_{1}e_{1}-z_{2}%
e_{2}-z_{3}e_{3}=x^{\dagger}+y^{\dagger}i,
\end{equation}
which was heavily used in the construction of $SL(2,C)$ instantons \cite{Ann}
in contrast to the complex conjugation%
\begin{equation}
z^{\ast}=z_{\mu}^{\ast}e_{\mu}=z_{0}^{\ast}e_{0}+z_{1}^{\ast}e_{1}+z_{2}%
^{\ast}e_{2}+z_{3}^{\ast}e_{3}=x-yi.
\end{equation}
The norm square of a biquarternion is defined to be%
\begin{equation}
|z|_{c}^{2}=z^{\circledast}z=(z_{0})^{2}+(z_{1})^{2}+(z_{2})^{2}+(z_{3})^{2},
\end{equation}
which is a \textit{complex} number in general as a subscript $c$ is used in
the norm.

We now review the biquaternion construction of $SL(2,C)$ instantons which
extends the quaternion construction of ADHM $SU(2)$ instantons. The first step
was to introduce the $(k+1)\times k$ biquarternion matrix $\Delta(x)=a+bx$%

\begin{equation}
\Delta(x)_{ab}=a_{ab}+b_{ab}x,\text{ }a_{ab}=a_{ab}^{\mu}e_{\mu},b_{ab}%
=b_{ab}^{\mu}e_{\mu} \label{ab}%
\end{equation}
where $a_{ab}^{\mu}$ and $b_{ab}^{\mu}$ are complex numbers, and $a_{ab}$ and
$b_{ab}$ are biquarternions. The biconjugation of the $\Delta(x)$ matrix is
defined to be%
\begin{equation}
\Delta(x)_{ab}^{\circledast}=\Delta(x)_{ba}^{\mu}e_{\mu}^{\dagger}%
=\Delta(x)_{ba}^{0}e_{0}-\Delta(x)_{ba}^{1}e_{1}-\Delta(x)_{ba}^{2}%
e_{2}-\Delta(x)_{ba}^{3}e_{3}.
\end{equation}
The quadratic condition of $SL(2,C)$ instantons reads%

\begin{equation}
\Delta(x)^{\circledast}\Delta(x)=f^{-1}=\text{symmetric, non-singular }k\times
k\text{ matrix for }x\notin J\text{,} \label{ff}%
\end{equation}
from which\ we can deduce that $a^{\circledast}a,b^{\circledast}%
a,a^{\circledast}b$ and $b^{\circledast}b$ are all symmetric matrices. The
choice of \textit{biconjugation} operation was crucial for the construction of
the $SL(2,C)$ instantons. On the other hand, for $x\in J,$ $\det
\Delta(x)^{\circledast}\Delta(x)=0$. The set $J$ is called singular locus or
"jumping lines". There are no jumping lines for the case of $SU(2)$ instantons
on $S^{4}$. In the $Sp(1)$ quaternion case, the symmetric condition on
$f^{-1}$ implies $f^{-1}$ is real; while for the $SL(2,C)$ biquaternion case,
it implies $f^{-1}$ is \textit{complex} which means $[\Delta(x)^{\circledast
}\Delta(x)]_{ij}^{\mu}=0$ for $\mu=1,2,3.$

To construct the self-dual gauge field, we introduce a $(k+1)\times1$
dimensional biquaternion vector $v(x)$ satisfying the following two conditions%
\begin{subequations}
\begin{align}
v^{\circledast}(x)\Delta(x)  &  =0,\label{null}\\
v^{\circledast}(x)v(x)  &  =1 \label{norm2}%
\end{align}
where $v(x)$ is fixed up to a $SL(2,C)$ gauge transformation%
\end{subequations}
\begin{equation}
v(x)\longrightarrow v(x)g(x),\text{ \ \ }g(x)\in\text{ }1\times1\text{
Biquaternion}.
\end{equation}
Note that in general a $SL(2,C)$ matrix can be written in terms of a
$1\times1$ biquaternion as%
\begin{equation}
g=\frac{q_{\mu}e_{\mu}}{\sqrt{q^{\circledast}q}}=\frac{q_{\mu}e_{\mu}}%
{|q|_{c}}.
\end{equation}
The next step is to define the gauge field%

\begin{equation}
G_{\mu}(x)=v^{\circledast}(x)\partial_{\mu}v(x), \label{A}%
\end{equation}
which is a $1\times1$ biquaternion. The $SL(2,C)$ gauge transformation of the
gauge field is%

\begin{align}
G_{\mu}(x)-  &  >G^{\prime}(x)=(g^{\circledast}(x)v^{\circledast}%
(x))\partial_{\mu}(v(x)g(x))\nonumber\\
&  =g^{\circledast}(x)G_{\mu}(x)g(x)+g^{\circledast}(x)\partial_{\mu}g(x)
\end{align}
where in the calculation Eq.(\ref{norm2}) has been used. Note that, unlike the
case for $Sp(1)$, $G_{\mu}(x)$ needs not to be anti-Hermitian.

One can then define the $SL(2,C)$ field strength
\begin{equation}
F_{\mu\nu}=\partial_{\mu}G_{\nu}(x)+G_{\mu}(x)G_{\nu}(x)-[\mu
\longleftrightarrow\nu],
\end{equation}
and prove the self-duality of $F_{\mu\nu}$. To count the number of moduli
parameters for the $SL(2,C)$ $k$-instantons, one can use transformations which
preserve conditions Eq.(\ref{ff}), Eq.(\ref{null}) and Eq.(\ref{norm2}), and
the definition of $G_{\mu}$ in Eq.(\ref{A}) to bring $a$ and $b$ in
Eq.(\ref{ab}) into the following simple canonical form%
\begin{equation}
b=%
\begin{bmatrix}
0_{1\times k}\\
I_{k\times k}%
\end{bmatrix}
,a=%
\begin{bmatrix}
\lambda_{1\times k}\\
-y_{k\times k}%
\end{bmatrix}
\label{ab2}%
\end{equation}
where $\lambda$ and $y$ are biquaternion matrices with orders $1\times k$ and
\ $k\times k$ respectively, and $y$ is symmetric%

\begin{equation}
y=y^{T}. \label{y}%
\end{equation}
Thus the constraints for the moduli parameters are%
\begin{equation}
a_{ci}^{\circledast}a_{cj}=0,i\neq j,\text{ and \ }y_{ij}=y_{ji}. \label{dof}%
\end{equation}
The total number of moduli parameters for $k$-instanton can be calculated
through Eq.(\ref{dof}) to be%
\begin{equation}
\text{\# of moduli for }SL(2,C)\text{ }k\text{-instantons}=16k-6,
\end{equation}
which is twice of that of the case of $Sp(1).$ Roughly speaking, there are
$8k$ parameters for instanton "biquaternion positions" and $8k$ parameters for
instanton "sizes". Finally one has to subtract an overall $SL(2,C)$ gauge
group degree of freedom $6.$

We provide two explicit examples of $SL(2,C)$ instantons here. These will be
used in secion IV for the discussion of instanton sheaves.

\subsection{The $SL(2,C)$ CFTW $k$-instantons}

We choose the biquaternion $\lambda_{j}$ in Eq.(\ref{ab2}) to be $\lambda
_{j}e_{0}$ with $\lambda_{j}$ a \textit{complex} number, and choose
$y_{ij}=y_{j}\delta_{ij}$ to be a diagonal matrix with $y_{j}=y_{j\mu}e_{\mu}$
a biquaternion. That is%

\begin{equation}
\Delta(x)=%
\begin{bmatrix}
\lambda_{1} & \lambda_{2} & ... & \lambda_{k}\\
x-y_{1} & 0 & ... & 0\\
0 & x-y_{2} & ... & 0\\
. & ... & ... & ...\\
0 & 0 & ... & x-y_{k}%
\end{bmatrix}
, \label{delta}%
\end{equation}
which satisfies the constraint in Eq.(\ref{dof}). Let%
\begin{equation}
v=\frac{1}{\sqrt{\phi}}%
\begin{bmatrix}
1\\
-q_{1}\\
.\\
-q_{k}%
\end{bmatrix}
,
\end{equation}
then
\begin{equation}
q_{j}=\frac{\lambda_{j}(x_{\mu}-y_{j\mu})e_{\mu}}{|x-y_{j}|_{c}^{2}%
},j=1,2,...,k,
\end{equation}
and%

\begin{equation}
v=\frac{1}{\sqrt{\phi}}%
\begin{bmatrix}
1\\
-\frac{\lambda_{1}(x_{\mu}-y_{1\mu})e_{\mu}}{|x-y_{1}|_{c}^{2}}\\
.\\
-\frac{\lambda_{k}(x_{\mu}-y_{k\mu})e_{\mu}}{|x-y_{k}|_{c}^{2}}%
\end{bmatrix}
\end{equation}
with

\bigskip%
\begin{equation}
\phi=1+\frac{\lambda_{1}\lambda_{1}^{\circledast}}{|x-y_{1}|_{c}^{2}%
}+...+\frac{\lambda_{k}\lambda_{k}^{\circledast}}{|x-y_{k}|_{c}^{2}}%
\end{equation}
where $\phi$ is a complex-valued function in general. One can calculate the
gauge potential as%
\begin{align}
G_{\mu}  &  =v^{\circledast}\partial_{\mu}v=\frac{1}{4}[e_{\mu}^{\dagger
}e_{\nu}-e_{\nu}^{\dagger}e_{\mu}]\partial_{\nu}\ln(1+\frac{\lambda_{1}^{2}%
}{|x-y_{1}|^{2}}+...+\frac{\lambda_{k}^{2}}{|x-y_{k}|^{2}})\nonumber\\
&  =\frac{1}{4}[e_{\mu}^{\dagger}e_{\nu}-e_{\nu}^{\dagger}e_{\mu}%
]\partial_{\nu}\ln(\phi).
\end{align}

To get non-removable singularities, one needs to calculate zeros of%
\begin{equation}
\phi=1+\frac{\lambda_{1}\lambda_{1}^{\circledast}}{|x-y_{1}|_{c}^{2}%
}+...+\frac{\lambda_{k}\lambda_{k}^{\circledast}}{|x-y_{k}|_{c}^{2}},
\end{equation}
or%
\begin{equation}
|x-y_{1}|_{c}^{2}|x-y_{2}|_{c}^{2}\cdot\cdot\cdot|x-y_{k}|_{c}^{2}\phi
=P_{2k}(x)+iP_{2k-1}(x)=0. \label{jump-k}%
\end{equation}
For the $SL(2,C)$ CFTW $k$-instanton case, one encounters intersections of
zeros of $P_{2k}(x)$ and $P_{2k-1}(x)$ polynomials with degrees $2k$ and
$2k-1$ respectively%
\begin{equation}
P_{2k}(x)=0,\text{ \ }P_{2k-1}(x)=0.
\end{equation}

\subsection{The General $SL(2,C)$ $2$-instanton Solutions}

For this case we choose the following $\Delta(x)$ matrix with $y_{12}=y_{21}$%

\begin{equation}
\Delta(x)=%
\begin{bmatrix}
\lambda_{1} & \lambda_{2}\\
x-y_{1} & -y_{12}\\
-y_{21} & x-y_{2}%
\end{bmatrix}
, \label{y12}%
\end{equation}

\begin{equation}
\Delta^{\circledast}(x)=%
\begin{bmatrix}
\lambda_{1}^{\circledast} & x^{\circledast}-y_{1}^{\circledast} &
-y_{12}^{\circledast}\\
\lambda_{2}^{\circledast} & -y_{12}^{\circledast} & x^{\circledast}%
-y_{2}^{\circledast}%
\end{bmatrix}
.
\end{equation}
The condition on $\Delta^{\circledast}(x)\Delta(x)$%
\begin{equation}
\Delta^{\circledast}(x)\Delta(x)=%
\begin{bmatrix}
\lambda_{1}^{\circledast}\lambda_{1}+(x^{\circledast}-y_{1}^{\circledast
})(x-y_{1})+y_{12}^{\circledast}y_{12} & \lambda_{1}^{\circledast}\lambda
_{2}-(x^{\circledast}-y_{1}^{\circledast})y_{12}-y_{12}^{\circledast}%
(x-y_{2})\\
\lambda_{2}^{\circledast}\lambda_{1}-y_{12}^{\circledast}(x-y_{1}%
)-(x^{\circledast}-y_{2}^{\circledast})y_{12} & \lambda_{2}^{\circledast
}\lambda_{2}+y_{12}^{\circledast}y_{12}+(x^{\circledast}-y_{2}^{\circledast
})(x-y_{2})
\end{bmatrix}
\label{2-instan}%
\end{equation}
in Eq.(\ref{ff}) is\bigskip%

\begin{equation}
\lambda_{2}^{\circledast}\lambda_{1}-\lambda_{1}^{\circledast}\lambda
_{2}=y_{12}^{\circledast}(y_{2}-y_{1})+(y_{1}^{\circledast}-y_{2}%
^{\circledast})y_{12},
\end{equation}
which is linear in the biquaternion $y_{12}$ instead of a quadratic equation,
and $y_{12}$ can be easily solved to be%
\begin{equation}
y_{12}=\frac{1}{2}\frac{(y_{1}-y_{2})}{|y_{1}-y_{2}|_{c}^{2}}(\lambda
_{2}^{\circledast}\lambda_{1}-\lambda_{1}^{\circledast}\lambda_{2}).
\label{y123}%
\end{equation}
The number of moduli for $SL(2,C)$ $2$-instanton solutions is $26$ as
expected. This general $2$-instanton solutions contain the previous CFTW
$2$-instanton solutions as a subset. We will see that although the vector
bundle description on $CP^{3}$ remains valid for the case of $SL(2,C)$ CFTW
$2$-instanton solutions, for some points on $CP^{3}$ and some subset of the
$26D$ moduli space of the general $SL(2,C)$ $2$-instanton solutions, the
vector bundle description breaks down, and one is led to use sheaf description
for these non-compact YM instantons or " instanton sheaves" on $CP^{3}%
$\cite{math2}.

For the singularities of general $2$-instanton solutions, one needs to
calculate zeros of the determinant%
\begin{align}
\det\Delta_{2-ins}(x)^{\circledast}\Delta_{2-ins}(x)  &  =|x-y_{1}|_{c}%
^{2}|x-y_{2}|_{c}^{2}+|\lambda_{2}|_{c}^{2}|x-y_{1}|_{c}^{2}+|\lambda_{1}%
|_{c}^{2}|x-y_{2}|_{c}^{2}\nonumber\\
&  +y_{12}^{\circledast}(x-y_{1})y_{12}^{\circledast}(x-y_{2})+(x-y_{2}%
)^{\circledast}y_{12}(x-y_{1})^{\circledast}y_{12}\nonumber\\
&  -y_{12}^{\circledast}(x-y_{1})\lambda_{1}^{\circledast}\lambda_{2}%
-\lambda_{2}^{\circledast}\lambda_{1}(x-y_{1})^{\circledast}y_{12}\nonumber\\
&  -(x-y_{2})^{\circledast}y_{12}\lambda_{1}^{\circledast}\lambda_{2}%
-\lambda_{2}^{\circledast}\lambda_{1}y_{12}^{\circledast}(x-y_{2})\nonumber\\
&  +|y_{12}|_{c}^{2}(|\lambda_{2}|_{c}^{2}+|\lambda_{1}|_{c}^{2})+|y_{12}%
|_{c}^{4}\nonumber\\
&  =0 \label{jump-2}%
\end{align}
where $y_{12}$ is given by Eq.(\ref{y123}). In calculating the determinant,
one notices that $\Delta(x)^{\circledast}\Delta(x)$ in Eq.(\ref{2-instan}) is
a symmetric matrix with complex number entries. So there is no ambiguity in
the determinant calculation.

\section{Solutions of Complex ADHM Equations and Monad Construction}

In this section, we will show that the $SL(2,C)$ YM instanton solutions
constructed in \cite{Ann} are solutions of the complex version of the ADHM
equations \cite{Donald}. Along with the calculation, we will first identify
the \textit{complex ADHM data }$(B_{lm},I_{m,}J_{m})$ with $l,m=1,2.$ We then
identify the corresponding $\alpha$ and $\beta$ matrices in the monad
construction on $CP^{3}$. \bigskip These identifications will enable us to
calculate in the next section the existence of points on $CP^{3}$ with some
instanton moduli where the vector bundle description of $SL(2,C)$
$2$-instanton on $CP^{3}$ breaks down, and one is led to use sheaf description
for $SL(2,C)$ non-compact YM instantons.

\subsection{The Complex ADHM Equations}

To do the calculation, we will need the explicit matrix representation (EMR)
of the biquaternion\bigskip%
\begin{equation}
e_{0}\rightarrow%
\begin{bmatrix}
1 & 0\\
0 & 1
\end{bmatrix}
,e_{1}\rightarrow-i\sigma_{1}=%
\begin{bmatrix}
0 & -i\\
-i & 0
\end{bmatrix}
,e_{2}\rightarrow-i\sigma_{2}=%
\begin{bmatrix}
0 & -1\\
1 & 0
\end{bmatrix}
,e_{3}\rightarrow-i\sigma_{3}=%
\begin{bmatrix}
-i & 0\\
0 & i
\end{bmatrix}
.
\end{equation}
So in the EMR, a biquaternion can be written as a $2\times2$ complex matrix
\begin{align}
z  &  =z^{0}e_{0}+z^{1}e_{1}+z^{2}e_{2}+z^{3}e_{3}\nonumber\\
&  =%
\begin{bmatrix}
\left(  a^{0}+b^{3}\right)  +i\left(  b^{0}-a^{3}\right)  & \left(
-a^{2}+b^{1}\right)  +i\left(  -b^{2}-a^{1}\right) \\
\left(  a^{2}+b^{1}\right)  +i\left(  b^{2}-a^{1}\right)  & \left(
a^{0}-b^{3}\right)  +i\left(  b^{0}+a^{3}\right)
\end{bmatrix}
\end{align}
where $a^{\mu}$ and $b^{\mu}$ are real and imaginary parts of $z^{\mu}$
respectively. The biconjugation is%
\begin{align}
z^{\circledast}  &  =z_{0}e_{0}-z_{1}e_{1}-z_{2}e_{2}-z_{3}e_{3}\nonumber\\
&  =%
\begin{bmatrix}
\left(  a^{0}-b^{3}\right)  +i\left(  b^{0}+a^{3}\right)  & \left(
a^{2}-b^{1}\right)  +i\left(  b^{2}+a^{1}\right) \\
\left(  -a^{2}-b^{1}\right)  +i\left(  -b^{2}+a^{1}\right)  & \left(
a^{0}+b^{3}\right)  +i\left(  b^{0}-a^{3}\right)
\end{bmatrix}
.
\end{align}
The norm square of a biquarternion used in this paper is defined to be%
\begin{align}
z^{\circledast}z  &  =zz^{\circledast}\nonumber\\
&  =%
\begin{bmatrix}
z^{0}+iz^{3} & z^{2}+iz^{1}\\
-\left(  z^{2}-iz^{1}\right)  & z^{0}-iz^{3}%
\end{bmatrix}%
\begin{bmatrix}
z^{0}-iz^{3} & -\left(  z^{2}+iz^{1}\right) \\
z^{2}-iz^{1} & z^{0}+iz^{3}%
\end{bmatrix}
\nonumber\\
&  =%
\begin{bmatrix}
\left(  z^{0}\right)  ^{2}+\left(  z^{1}\right)  ^{2}+\left(  z^{2}\right)
^{2}+\left(  z^{3}\right)  ^{2} & 0\\
0 & \left(  z^{0}\right)  ^{2}+\left(  z^{1}\right)  ^{2}+\left(
z^{2}\right)  ^{2}+\left(  z^{3}\right)  ^{2}%
\end{bmatrix}
.
\end{align}

We are now ready to show that the $SL(2,C)$ YM instanton solutions constructed
using biquaternion in the last section are indeed solutions of complex version
of the ADHM equations. We will need to first identify the \textit{complex ADHM
data} $(B_{lm},I_{m,}J_{m})$ with $l,m=1,2.$ For simplicity, we will do the
calculation in the canonical form in Eq.(\ref{ab2}) and Eq.(\ref{y}) with
constraints on the moduli parameters in Eq.(\ref{dof}). For the $k$-instanton
case, the EMR of\ the $(k+1)\times k$ biquaternion matrix $a$ in
Eq.(\ref{ab2})
\begin{equation}
a=%
\begin{bmatrix}
\lambda_{1} & \lambda_{2} & ... & \lambda_{k}\\
y_{11} & y_{12} & ... & y_{1k}\\
y_{21} & y_{22} & ... & y_{2k}\\
.. & ... & ... & ...\\
y_{k1} & y_{k2} & ... & y_{kk}%
\end{bmatrix}
\end{equation}
with $y_{ij}=y_{ji}$ can be written as a $2(k+1)\times2k$ complex matrix
\begin{equation}
a=%
\begin{bmatrix}
\lambda_{1}^{0}-i\lambda_{1}^{3} & -\left(  \lambda_{1}^{2}+i\lambda_{1}%
^{1}\right)  & \lambda_{2}^{0}-i\lambda_{2}^{3} & -\left(  \lambda_{2}%
^{2}+i\lambda_{2}^{1}\right)  & ... & \lambda_{k}^{0}-i\lambda_{k}^{3} &
-\left(  \lambda_{k}^{2}+i\lambda_{k}^{1}\right) \\
\lambda_{1}^{2}-i\lambda_{1}^{1} & \lambda_{1}^{0}+i\lambda_{1}^{3} &
\lambda_{2}^{2}-i\lambda_{2}^{1} & \lambda_{2}^{0}+i\lambda_{2}^{3} & .. &
\lambda_{k}^{2}-i\lambda_{k}^{1} & \lambda_{k}^{0}+i\lambda_{k}^{3}\\
y_{11}^{0}-iy_{11}^{3} & -\left(  y_{11}^{2}+iy_{11}^{1}\right)  & y_{12}%
^{0}-iy_{12}^{3} & -\left(  y_{12}^{2}+iy_{12}^{1}\right)  & ... & y_{1k}%
^{0}-iy_{1k}^{3} & -\left(  y_{1k}^{2}+iy_{1k}^{1}\right) \\
y_{11}^{2}-iy_{11}^{1} & y_{11}^{0}+iy_{11}^{3} & y_{12}^{2}-iy_{12}^{1} &
y_{12}^{0}+iy_{12}^{3} & ... & y_{1k}^{2}-iy_{1k}^{1} & y_{1k}^{0}+iy_{1k}%
^{3}\\
y_{12}^{0}-iy_{12}^{3} & -\left(  y_{12}^{2}+iy_{12}^{1}\right)  & y_{22}%
^{0}-iy_{22}^{3} & -\left(  y_{22}^{2}+iy_{22}^{1}\right)  & ... & y_{2k}%
^{0}-iy_{2k}^{3} & -\left(  y_{2k}^{2}+iy_{2k}^{1}\right) \\
y_{12}^{2}-iy_{12}^{1} & y_{12}^{0}+iy_{12}^{3} & y_{22}^{2}-iy_{22}^{1} &
y_{22}^{0}+iy_{22}^{3} & ... & y_{2k}^{2}-iy_{2k}^{1} & y_{2k}^{2}-iy_{2k}%
^{1}\\
.. & ... & ... & ... & ... & ... & ...\\
y_{1k}^{0}-iy_{1k}^{3} & -\left(  y_{1k}^{2}+iy_{1k}^{1}\right)  & y_{2k}%
^{0}-iy_{2k}^{3} & -\left(  y_{2k}^{2}+iy_{2k}^{1}\right)  & ... & y_{kk}%
^{0}-iy_{2k}^{3} & -\left(  y_{kk}^{2}+iy_{kk}^{1}\right) \\
y_{1k}^{2}-iy_{1k}^{1} & y_{1k}^{0}+iy_{1k}^{3} & y_{2k}^{2}-iy_{2k}^{1} &
y_{2k}^{0}+iy_{2k}^{3} & ... & y_{kk}^{2}-iy_{kk}^{1} & y_{kk}^{0}+iy_{kk}^{3}%
\end{bmatrix}
.
\end{equation}
where $\lambda_{j}^{i}$ and $y_{jk}^{i}$ are all complex numbers. We then do
the following rearrangement and identification for the complex ADHM data
\begin{align}
a  &  \rightarrow%
\begin{bmatrix}
\lambda_{1}^{0}-i\lambda_{1}^{3} & \lambda_{2}^{0}-i\lambda_{2}^{3} & .. &
\lambda_{k}^{0}-i\lambda_{k}^{3} & -\left(  \lambda_{1}^{2}+i\lambda_{1}%
^{1}\right)  & -\left(  \lambda_{2}^{2}+i\lambda_{2}^{1}\right)  & ... &
-\left(  \lambda_{k}^{2}+i\lambda_{k}^{1}\right) \\
\lambda_{1}^{2}-i\lambda_{1}^{1} & \lambda_{2}^{2}-i\lambda_{2}^{1} & .. &
\lambda_{k}^{2}-i\lambda_{k}^{1} & \lambda_{1}^{0}+i\lambda_{1}^{3} &
\lambda_{2}^{0}+i\lambda_{2}^{3} & ... & \lambda_{k}^{0}+i\lambda_{k}^{3}\\
y_{11}^{0}-iy_{11}^{3} & y_{12}^{0}-iy_{12}^{3} & ... & y_{1k}^{0}-iy_{1k}^{3}
& -\left(  y_{11}^{2}+iy_{11}^{1}\right)  & -\left(  y_{12}^{2}+iy_{12}%
^{1}\right)  & ... & -\left(  y_{1k}^{2}+iy_{1k}^{1}\right) \\
y_{12}^{0}-iy_{12}^{3} & y_{22}^{0}-iy_{22}^{3} & ... & y_{2k}^{0}-iy_{2k}^{3}
& -\left(  y_{12}^{2}+iy_{12}^{1}\right)  & -\left(  y_{22}^{2}+iy_{22}%
^{1}\right)  & ... & -\left(  y_{2k}^{2}+iy_{2k}^{1}\right) \\
.. & ... & ... & ... & ... & ... & ... & ...\\
y_{1k}^{0}-iy_{1k}^{3} & y_{2k}^{0}-iy_{2k}^{3} & ... & y_{kk}^{0}-iy_{kk}^{3}
& -\left(  y_{1k}^{2}+iy_{1k}^{1}\right)  & -\left(  y_{2k}^{2}+iy_{2k}%
^{1}\right)  & ... & -\left(  y_{kk}^{2}+iy_{kk}^{1}\right) \\
y_{11}^{2}-iy_{11}^{1} & y_{12}^{2}-iy_{12}^{1} & ... & y_{1k}^{2}-iy_{1k}^{1}
& y_{11}^{0}+iy_{11}^{3} & y_{12}^{0}+iy_{12}^{3} & ... & y_{1k}^{0}%
+iy_{1k}^{3}\\
y_{12}^{2}-iy_{12}^{1} & y_{22}^{2}-iy_{22}^{1} & ... & y_{2k}^{2}-iy_{2k}^{1}
& y_{12}^{0}+iy_{12}^{3} & y_{22}^{0}+iy_{22}^{3} & ... & y_{2k}^{0}%
+iy_{2k}^{3}\\
.. & ... & ... & ... & ... & ... & ... & ...\\
y_{1k}^{2}-iy_{1k}^{1} & y_{2k}^{2}-iy_{2k}^{1} & ... & y_{kk}^{2}-iy_{kk}^{1}
& y_{1k}^{0}+iy_{1k}^{3} & y_{2k}^{0}+iy_{2k}^{3} & ... & y_{kk}^{0}%
+iy_{kk}^{3}%
\end{bmatrix}
\\
&  =%
\begin{bmatrix}
J_{1} & J_{2}\\
B_{11} & B_{21}\\
B_{12} & B_{22}%
\end{bmatrix}
\label{id}%
\end{align}
where we have done the \textit{rearrangement rule }for an element $z_{ij}$ in
$a$\textit{ }
\begin{align}
z_{2n-1,2m-1}  &  \rightarrow z_{n,m}\text{ },\nonumber\\
z_{2n-1,2m}  &  \rightarrow z_{n,k+m}\text{ },\nonumber\\
z_{2n,2m-1}  &  \rightarrow z_{k+n,m}\text{ },\nonumber\\
z_{2n,2m}  &  \rightarrow z_{k+n,k+m}.
\end{align}
A simple example for the case of $k=2$ will be given in Eq.(\ref{id2}). In
Eq.(\ref{id}) $B_{ij}$ are $k\times k$ complex matrices and $J_{i}$ are
$2\times k$ complex matrices. Similarly for $a^{\circledast}$ we get%
\begin{align}
a^{\circledast}  &  =%
\begin{bmatrix}
\lambda_{1}^{\circledast} & y_{11}^{\circledast} & y_{12}^{\circledast} & .. &
y_{1k}^{\circledast}\\
\lambda_{2}^{\circledast} & y_{12}^{\circledast} & y_{22}^{\circledast} & .. &
y_{2k}^{\circledast}\\
.. & ... & ... & ... & ...\\
\lambda_{k}^{\circledast} & y_{1k}^{\circledast} & y_{2k}^{\circledast} & .. &
y_{kk}^{\circledast}%
\end{bmatrix}
\rightarrow\\
&
\begin{bmatrix}
\lambda_{1}^{0}+i\lambda_{1}^{3} & \lambda_{1}^{2}+i\lambda_{1}^{1} &
y_{11}^{0}+iy_{11}^{3} & y_{12}^{0}+iy_{12}^{3} & ... & y_{1k}^{0}+iy_{1k}^{3}
& y_{11}^{2}+iy_{11}^{1} & y_{12}^{2}+iy_{12}^{1} & ... & y_{1k}^{2}%
+iy_{1k}^{1}\\
\lambda_{2}^{0}+i\lambda_{2}^{3} & \lambda_{2}^{2}+i\lambda_{2}^{1} &
y_{12}^{0}+iy_{12}^{3} & y_{22}^{0}+iy_{22}^{3} & ... & y_{2k}^{0}+iy_{2k}^{3}
& y_{12}^{2}+iy_{12}^{1} & y_{22}^{2}+iy_{22}^{1} & ... & y_{2k}^{2}%
+iy_{2k}^{1}\\
.. & ... & ... & ... & ... & ... & ... & ... & ... & ...\\
\lambda_{k}^{0}+i\lambda_{k}^{3} & \lambda_{k}^{2}+i\lambda_{k}^{1} &
y_{1k}^{0}+iy_{1k}^{3} & y_{2k}^{0}+iy_{2k}^{3} & ... & y_{kk}^{0}+iy_{kk}^{3}
& y_{1k}^{2}+iy_{1k}^{1} & y_{2k}^{2}+iy_{2k}^{1} & ... & y_{kk}^{2}%
+iy_{kk}^{1}\\
-\lambda_{1}^{2}+i\lambda_{1}^{1} & \lambda_{1}^{0}-i\lambda_{1}^{3} &
-y_{11}^{2}+iy_{11}^{1} & -y_{12}^{2}+iy_{12}^{1} & ... & -y_{1k}^{2}%
+iy_{1k}^{1} & y_{11}^{0}-iy_{11}^{3} & y_{12}^{0}-iy_{12}^{3} & ... &
y_{1k}^{0}-iy_{1k}^{3}\\
-\lambda_{2}^{2}+i\lambda_{2}^{1} & \lambda_{2}^{0}-i\lambda_{2}^{3} &
-y_{12}^{2}+iy_{12}^{1} & -y_{22}^{2}+iy_{22}^{1} & ... & -y_{2k}^{2}%
+iy_{2k}^{1} & y_{12}^{0}-iy_{12}^{3} & y_{22}^{0}-iy_{22}^{3} & ... &
y_{2k}^{0}-iy_{2k}^{3}\\
.. & ... & ... & ... & ... & ... & ... & ... & ... & ...\\
-\lambda_{k}^{2}+i\lambda_{k}^{1} & \lambda_{k}^{0}-i\lambda_{k}^{3} &
-y_{1k}^{2}+iy_{1k}^{1} & -y_{2k}^{2}+iy_{2k}^{1} & ... & -y_{kk}^{2}%
+iy_{kk}^{1} & y_{1k}^{0}-iy_{1k}^{3} & y_{2k}^{0}-iy_{2k}^{3} & ... &
y_{kk}^{0}-iy_{kk}^{3}%
\end{bmatrix}
\\
&  =%
\begin{bmatrix}
-I_{2} & B_{22} & -B_{21}\\
I_{1} & -B_{12} & B_{11}%
\end{bmatrix}
\end{align}
where $I_{j}$ are $k\times2$ matrices. The next step is to impose the
conditions in Eq.(\ref{dof}). The EMR of the biquaternion matrix
$a^{\circledast}a$%
\begin{equation}
\left(  a^{\circledast}a\right)  =%
\begin{bmatrix}
\left(  a^{\circledast}a\right)  _{11}^{0}e_{0} & 0 & ... & 0\\
0 & \left(  a^{\circledast}a\right)  _{22}^{0}e_{0} & ... & 0\\
.. & ... & ... & ...\\
0 & 0 & ... & \left(  a^{\circledast}a\right)  _{kk}^{0}e_{0}%
\end{bmatrix}
\end{equation}
can be written as%
\begin{equation}
\left(  a^{\circledast}a\right)  =%
\begin{bmatrix}
\left(  a^{\circledast}a\right)  _{11}^{0} & 0 & 0 & 0 & ... & 0 & 0\\
0 & \left(  a^{\circledast}a\right)  _{11}^{0} & 0 & 0 & ... & 0 & 0\\
0 & 0 & \left(  a^{\circledast}a\right)  _{22}^{0} & 0 & ... & 0 & 0\\
0 & 0 & 0 & \left(  a^{\circledast}a\right)  _{22}^{0} & ... & 0 & 0\\
.. & ... & ... & ... & ... & ... & ...\\
0 & 0 & 0 & 0 & ... & \left(  a^{\circledast}a\right)  _{kk}^{0} & 0\\
0 & 0 & 0 & 0 & ... & 0 & \left(  a^{\circledast}a\right)  _{kk}^{0}%
\end{bmatrix}
.
\end{equation}
After the\ rearrangement and identification, we get
\begin{align}
a^{\circledast}a  &  \rightarrow%
\begin{bmatrix}
\left(  a^{\circledast}a\right)  _{11}^{0} & 0 & ... & 0 & 0 & 0 & 0 & 0\\
0 & \left(  a^{\circledast}a\right)  _{22}^{0} & ... & 0 & 0 & 0 & 0 & 0\\
.. & ... & ... & ... & 0 & 0 & 0 & 0\\
0 & 0 & ... & \left(  a^{\circledast}a\right)  _{kk}^{0} & 0 & 0 & 0 & 0\\
0 & 0 & 0 & 0 & \left(  a^{\circledast}a\right)  _{11}^{0} & 0 & ... & 0\\
0 & 0 & 0 & 0 & 0 & \left(  a^{\circledast}a\right)  _{22}^{0} & ... & 0\\
0 & 0 & 0 & 0 & ... & ... & ... & ...\\
0 & 0 & 0 & 0 & 0 & 0 & ... & \left(  a^{\circledast}a\right)  _{kk}^{0}%
\end{bmatrix}
\\
&  =%
\begin{bmatrix}
-I_{2}J_{1}+B_{22}B_{11}-B_{21}B_{12} & -I_{2}J_{2}+B_{22}B_{21}-B_{21}%
B_{22}\\
I_{1}J_{1}+B_{11}B_{12}-B_{12}B_{11} & I_{1}J_{2}+B_{11}B_{22}-B_{12}B_{21}%
\end{bmatrix}
,
\end{align}
which leads immediately to the complex ADHM equations
\begin{subequations}
\begin{align}
\left[  B_{11},B_{12}\right]  +I_{1}J_{1}  &  =0,\label{adhm1}\\
\left[  B_{21},B_{22}\right]  +I_{2}J_{2}  &  =0,\label{adhm2}\\
\left[  B_{11},B_{22}\right]  +\left[  B_{21},B_{12}\right]  +I_{1}J_{2}%
+I_{2}J_{1}  &  =0. \label{adhm3}%
\end{align}
For the case of $SU(2)$ ADHM instantons, we impose the conditions%
\end{subequations}
\begin{subequations}
\begin{align}
I_{1}  &  =J^{\dagger},I_{2}=-I,J_{1}=I^{\dagger},J_{2}=J,\nonumber\\
B_{11}  &  =B_{2}^{\dagger},B_{12}=B_{1}^{\dagger},B_{21}=-B_{1},B_{22}=B_{2}%
\end{align}
to get the real ADHM equations%
\end{subequations}
\begin{subequations}
\begin{align}
\left[  B_{1},B_{2}\right]  +IJ  &  =0,\\
\left[  B_{1},B_{1}^{\dagger}\right]  +\left[  B_{2},B_{2}^{\dagger}\right]
+II^{\dagger}-J^{\dagger}J  &  =0.
\end{align}
We thus complete the proof that the $SL(2,C)$ YM instanton solutions we
constructed using the biquaternion method in the last section are solutions of
the complex version of the ADHM equations.

\subsection{The Monad Construction}

In the rest of this section, we construct the $\bigskip\alpha\,$and $\beta$
matrices in the monad construction as functions of homogeneous coordinates
$z,w,x,y$ of $CP^{3}$ which will be used in the next section. We define%
\end{subequations}
\begin{subequations}
\begin{align}
\alpha &  =%
\begin{bmatrix}
zB_{11}+wB_{21}+x\\
zB_{12}+wB_{22}+y\\
zJ_{1}+wJ_{2}%
\end{bmatrix}
,\\
\beta &  =%
\begin{bmatrix}
-zB_{12}-wB_{22}-y & zB_{11}+wB_{21}+x & zI_{1}+wI_{2}%
\end{bmatrix}
.
\end{align}
Similar to the real ADHM equations, it can be shown that the condition%
\end{subequations}
\begin{equation}
\beta\alpha=0 \label{ker}%
\end{equation}
is satisfied if and only if the complex ADHM equations in Eq.(\ref{adhm1}) to
Eq.(\ref{adhm3}) holds.

In the monad construction of the holomorphic vector bundles, Eq.(\ref{ker})
implies Im $\alpha$ is a subspace of Ker $\beta$ which allows one to consider
the quotient vector space Ker $\beta$/ Im $\alpha$ at each point of $CP^{3}$.
If the map $\beta$ is surjective and the map $\alpha$ is injective, then
$\dim($Ker $\beta$/ Im $\alpha)=k+2-k=2$ on every points of $CP^{3}$, thus one
can use holomorphic vector bundles to describe instantons. This is the case of
$SU(2)$ ADHM instantons. For the case of $SL(2,C)$ instantons, either $\beta$
may not be surjective or $\alpha$ may not be injective at some points of
$CP^{3}$ for some ADHM data, the dimension of $($Ker $\beta$/ Im $\alpha)$ may
vary from point to point on $CP^{3}$, and one is led to use sheaf description
for these non-compact $SL(2,C)$ YM instantons or "instanton sheaves" on
$CP^{3}$\cite{math2}. These instanton sheaves will be discussed in the next section.

\section{The Yang-Mills $SL(2,C)$ $2$-instanton Sheaves}

As were shown in section II, there were jumping lines of the $SL(2,C)$ CFTW
$k$-Instanton solutions in Eq.(\ref{jump-k}) and the $SL(2,C)$ general
$2$-Instanton solutions in Eq.(\ref{jump-2}) on $S^{4}$ which can not be
gauged away as in the $SU(2)$ case. As a result, the vector bundle
descriptions on $CP^{3}$ for these cases may break down, and one is led to
introduce the sheaf structure on $CP^{3}$. In this section, we will apply
locally free conditions, or the costable and stable conditions introduced in
\cite{math2} to explicitly show that the vector bundle description of the
$SL(2,C)$ CFTW $k$-Instanton solutions on $CP^{3}$ remains valid, while that
of general $2$-instanton solutions breaks down.

\subsection{$SL(2,C)$ CFTW $k$-instanton Solutions are Locally Free}

For illustration, we calculate the $SL(2,C)$ CFTW $2$-instanton case. The
calculation of $k$-instanton case can be easily extended. For the
$2$-instanton case, $\lambda=\lambda^{0}e_{0}$ and $y$ is a digonal
biquaternion matrix%

\begin{align}
a  &  =%
\begin{bmatrix}
\lambda_{1} & \lambda_{2}\\
y_{11} & 0\\
0 & y_{22}%
\end{bmatrix}
=%
\begin{bmatrix}
p_{1}+iq_{1} & 0 & p_{2}+iq_{2} & 0\\
0 & p_{1}+iq_{1} & 0 & p_{2}+iq_{2}\\
y_{11}^{0}-iy_{11}^{3} & -\left(  y_{11}^{2}+iy_{11}^{1}\right)  & 0 & 0\\
y_{11}^{2}-iy_{11}^{1} & y_{11}^{0}+iy_{11}^{3} & 0 & 0\\
0 & 0 & y_{22}^{0}-iy_{22}^{3} & -\left(  y_{22}^{2}+iy_{22}^{1}\right) \\
0 & 0 & y_{22}^{2}-iy_{22}^{1} & y_{22}^{0}+iy_{22}^{3}%
\end{bmatrix}
\nonumber\\
&  \rightarrow%
\begin{bmatrix}
p_{1}+iq_{1} & p_{2}+iq_{2} & 0 & 0\\
0 & 0 & p_{1}+iq_{1} & p_{2}+iq_{2}\\
y_{11}^{0}-iy_{11}^{3} & 0 & -\left(  y_{11}^{2}+iy_{11}^{1}\right)  & 0\\
0 & y_{22}^{0}-iy_{22}^{3} & 0 & -\left(  y_{22}^{2}+iy_{22}^{1}\right) \\
y_{11}^{2}-iy_{11}^{1} & 0 & y_{11}^{0}+iy_{11}^{3} & 0\\
0 & y_{22}^{2}-iy_{22}^{1} & 0 & y_{22}^{0}+iy_{22}^{3}%
\end{bmatrix}
=%
\begin{bmatrix}
J_{1} & J_{2}\\
B_{11} & B_{21}\\
B_{12} & B_{22}%
\end{bmatrix}
\label{id2}%
\end{align}
where we have made a rearrangement for $a$ in the second line of
Eq.(\ref{id2}). Similarly we have%
\begin{align}
a^{\circledast}  &  =%
\begin{bmatrix}
\lambda_{1}^{\circledast} & y_{11}^{\circledast} & 0\\
\lambda_{2}^{\circledast} & 0 & y_{22}^{\circledast}%
\end{bmatrix}
\nonumber\\
&  =%
\begin{bmatrix}
p_{1}+iq_{1} & 0 & y_{11}^{0}+iy_{11}^{3} & y_{11}^{2}+iy_{11}^{1} & 0 & 0\\
0 & p_{1}+iq_{1} & -y_{11}^{2}+iy_{11}^{1} & y_{11}^{0}-iy_{11}^{3} & 0 & 0\\
p_{2}+iq_{2} & 0 & 0 & 0 & y_{22}^{0}+iy_{22}^{3} & y_{22}^{2}+iy_{22}^{1}\\
0 & p_{2}+iq_{2} & 0 & 0 & -y_{22}^{2}+iy_{22}^{1} & y_{22}^{0}-iy_{22}^{3}%
\end{bmatrix}
\nonumber\\
&  \rightarrow%
\begin{bmatrix}
p_{1}+iq_{1} & 0 & y_{11}^{0}+iy_{11}^{3} & 0 & y_{11}^{2}+iy_{11}^{1} & 0\\
p_{2}+iq_{2} & 0 & 0 & y_{22}^{0}+iy_{22}^{3} & 0 & y_{22}^{2}+iy_{22}^{1}\\
0 & p_{1}+iq_{1} & -y_{11}^{2}+iy_{11}^{1} & 0 & y_{11}^{0}-iy_{11}^{3} & 0\\
0 & p_{2}+iq_{2} & 0 & -y_{22}^{2}+iy_{22}^{1} & 0 & y_{22}^{0}-iy_{22}^{3}%
\end{bmatrix}
\nonumber\\
&  =%
\begin{bmatrix}
-I_{2} & B_{22} & -B_{21}\\
I_{1} & -B_{12} & B_{11}%
\end{bmatrix}
.
\end{align}
So we have the following identification%
\begin{subequations}
\begin{align}
I_{1}  &  =%
\begin{bmatrix}
0 & p_{1}+iq_{1}\\
0 & p_{2}+iq_{2}%
\end{bmatrix}
,I_{2}=%
\begin{bmatrix}
-\left(  p_{1}+iq_{1}\right)  & 0\\
-\left(  p_{2}+iq_{2}\right)  & 0
\end{bmatrix}
,J_{1}=%
\begin{bmatrix}
p_{1}+iq_{1} & p_{2}+iq_{2}\\
0 & 0
\end{bmatrix}
,J_{2}=%
\begin{bmatrix}
0 & 0\\
p_{1}+iq_{1} & p_{2}+iq_{2}%
\end{bmatrix}
,\\
B_{11}  &  =%
\begin{bmatrix}
y_{11}^{0}-iy_{11}^{3} & 0\\
0 & y_{22}^{0}-iy_{22}^{3}%
\end{bmatrix}
,B_{12}=%
\begin{bmatrix}
y_{11}^{2}-iy_{11}^{1} & 0\\
0 & y_{22}^{2}-iy_{22}^{1}%
\end{bmatrix}
,\\
B_{21}  &  =%
\begin{bmatrix}
-\left(  y_{11}^{2}+iy_{11}^{1}\right)  & 0\\
0 & -\left(  y_{22}^{2}+iy_{22}^{1}\right)
\end{bmatrix}
,B_{22}=%
\begin{bmatrix}
y_{11}^{0}+iy_{11}^{3} & 0\\
0 & y_{22}^{0}+iy_{22}^{3}%
\end{bmatrix}
.
\end{align}
It can be easily shown that for these parametrizations, the complex ADHM
equations in Eq.(\ref{adhm1}) to Eq.(\ref{adhm3}) are satisfied.

The next step is to check whether there exists common eigenvector $v$ in the
costable condition \cite{math2}%
\end{subequations}
\begin{subequations}
\begin{align}
\left(  zB_{11}+wB_{21}\right)  v &  =-xv,\label{a}\\
\left(  zB_{12}+wB_{22}\right)  v &  =-yv,\label{b}\\
\left(  zJ_{1}+wJ_{2}\right)  v &  =0.\label{c}%
\end{align}
If the common eigenvector $v$ exists, then the dimension of $($Ker $\beta$/ Im
$\alpha)$ discussed in the end of the last section will not be a constant. The
holomorphic vector bundle description on $CP^{3}$ will break down. From
Eq.(\ref{a}) and Eq.(\ref{b}), the possible solutions of $v$ are either $%
\begin{bmatrix}
1\\
0
\end{bmatrix}
$ or $%
\begin{bmatrix}
0\\
1
\end{bmatrix}
\,$, but these can not be the solution of Eq.(\ref{c}). So $SL(2,C)$ CFTW
$2$-instanton solutions are costable. Similar calculation can be done for the
case of stable condition. Thus $SL(2,C)$ CFTW $2$-instanton solutions are
locally free. Similar demonstration can be easily done for the $SL(2,C)$ CFTW
$k$-instanton solutions. We conclude that for the $SL(2,C)$ CFTW $k$-instanton
solutions with $10k$ moduli parameters, although the jumping lines exist on
$S^{4}$, the vector bundle description on $CP^{3}$ remains valid as in the
case of $SU(2)$ instantons.

\subsection{Breakdown of Vector Bundle Description}

In this subsection, we consider the case of complete known $SL(2,C)$
$2$-instanton solutions with $26$ moduli parameters. We will see that, for
some points on $CP^{3}$ and some subset of the $26D$ moduli space of
$2$-instanton solutions, the vector bundle description of $SL(2,C)$
$2$-instanton on $CP^{3}$ breaks down, and one is led to use sheaf description
for these non-compact YM instantons or "instanton sheaves" on $CP^{3}%
$\cite{math2}. Note that in \cite{LLT4} a duality symmetry among stability
conditions and costability conditions for YM instanton sheaf solutions was
pointed out with application to the known sheaf solutions. We thus will only
calculate the costability conditions in this paper. To proceed, we first
identify the ADHM data. let
\end{subequations}
\begin{equation}
\lambda_{1}=%
\begin{bmatrix}
\lambda_{1}^{0}-i\lambda_{1}^{3} & -\left(  \lambda_{1}^{2}+i\lambda_{1}%
^{1}\right)  \\
\lambda_{1}^{2}-i\lambda_{1}^{1} & \lambda_{1}^{0}+i\lambda_{1}^{3}%
\end{bmatrix}
,\lambda_{2}=%
\begin{bmatrix}
\lambda_{2}^{0}-i\lambda_{2}^{3} & -\left(  \lambda_{2}^{2}+i\lambda_{2}%
^{1}\right)  \\
\lambda_{2}^{2}-i\lambda_{2}^{1} & \lambda_{2}^{0}+i\lambda_{2}^{3}%
\end{bmatrix}
\end{equation}
and%
\begin{equation}
\lambda_{1}^{\circledast}=%
\begin{bmatrix}
\lambda_{1}^{0}+i\lambda_{1}^{3} & \lambda_{1}^{2}+i\lambda_{1}^{1}\\
-\left(  \lambda_{1}^{2}-i\lambda_{1}^{1}\right)   & \lambda_{1}^{0}%
-i\lambda_{1}^{3}%
\end{bmatrix}
,\lambda_{2}^{\circledast}=%
\begin{bmatrix}
\lambda_{2}^{0}+i\lambda_{2}^{3} & \lambda_{2}^{2}+i\lambda_{2}^{1}\\
-\left(  \lambda_{2}^{2}-i\lambda_{2}^{1}\right)   & \lambda_{2}^{0}%
-i\lambda_{2}^{3}%
\end{bmatrix}
.
\end{equation}
For further simplification, we define
\begin{subequations}
\begin{align}
l &  =%
\begin{vmatrix}
\lambda_{1}^{0} & \lambda_{1}^{3}\\
\lambda_{2}^{0} & \lambda_{2}^{3}%
\end{vmatrix}
-%
\begin{vmatrix}
\lambda_{1}^{1} & \lambda_{1}^{2}\\
\lambda_{2}^{1} & \lambda_{2}^{2}%
\end{vmatrix}
,\\
n &  =%
\begin{vmatrix}
\lambda_{1}^{0} & \lambda_{1}^{2}\\
\lambda_{2}^{0} & \lambda_{2}^{2}%
\end{vmatrix}
-%
\begin{vmatrix}
\lambda_{1}^{0} & \lambda_{1}^{1}\\
\lambda_{2}^{0} & \lambda_{2}^{1}%
\end{vmatrix}
,\\
m &  =%
\begin{vmatrix}
\lambda_{1}^{0} & \lambda_{1}^{1}\\
\lambda_{2}^{0} & \lambda_{2}^{1}%
\end{vmatrix}
-%
\begin{vmatrix}
\lambda_{1}^{2} & \lambda_{1}^{3}\\
\lambda_{2}^{2} & \lambda_{2}^{3}%
\end{vmatrix}
.
\end{align}
We choose the following biquaternions for $y_{11}$and $y_{22}$%
\end{subequations}
\begin{align}
z &  =z^{0}e_{0}+z^{1}e_{1}+z^{2}e_{2}+z^{3}e_{3}\nonumber\\
&  =%
\begin{bmatrix}
\left(  a^{0}+b^{3}\right)  +i\left(  b^{0}-a^{3}\right)   & \left(
-a^{2}+b^{1}\right)  +i\left(  -b^{2}-a^{1}\right)  \\
\left(  a^{2}+b^{1}\right)  +i\left(  b^{2}-a^{1}\right)   & \left(
a^{0}-b^{3}\right)  +i\left(  b^{0}+a^{3}\right)
\end{bmatrix}
\end{align}%
\begin{equation}
y_{11}=-de_{0}=%
\begin{bmatrix}
-d & 0\\
0 & -d
\end{bmatrix}
,y_{22}=de_{0}=%
\begin{bmatrix}
d & 0\\
0 & d
\end{bmatrix}
.
\end{equation}
One can then calculate $y_{12}$ by using Eq.(\ref{y123}) to get
\begin{equation}
y_{12}=\frac{-i}{2}%
\begin{bmatrix}
l & m+in\\
m+in & -l
\end{bmatrix}
.
\end{equation}
So we have the EMR of the biquaternion matrix
\begin{equation}%
\begin{bmatrix}
\lambda_{1} & \lambda_{2}\\
y_{11} & y_{12}\\
y_{12} & y_{22}%
\end{bmatrix}
=%
\begin{bmatrix}
\lambda_{1}^{0}-i\lambda_{1}^{3} & -\left(  \lambda_{1}^{2}+i\lambda_{1}%
^{1}\right)   & \lambda_{2}^{0}-i\lambda_{2}^{3} & -\left(  \lambda_{2}%
^{2}+i\lambda_{2}^{1}\right)  \\
\lambda_{1}^{2}-i\lambda_{1}^{1} & \lambda_{1}^{0}+i\lambda_{1}^{3} &
\lambda_{2}^{2}-i\lambda_{2}^{1} & \lambda_{2}^{0}+i\lambda_{2}^{3}\\
-d & 0 & \left(  \frac{-i}{2d}\right)  l & \left(  \frac{-i}{2d}\right)
\left(  m-in\right)  \\
0 & -d & \left(  \frac{-i}{2d}\right)  \left(  m+in\right)   & -\left(
\frac{-i}{2d}\right)  l\\
\left(  \frac{-i}{2d}\right)  l & \left(  \frac{-i}{2d}\right)  \left(
m-in\right)   & d & 0\\
\left(  \frac{-i}{2d}\right)  \left(  m+in\right)   & -\left(  \frac{-i}%
{2d}\right)  l & 0 & d
\end{bmatrix}
.
\end{equation}
After the rearrangement, we have the identification%
\begin{equation}%
\begin{bmatrix}
\lambda_{1}^{0}-i\lambda_{1}^{3} & \lambda_{2}^{0}-i\lambda_{2}^{3} & -\left(
\lambda_{1}^{2}+i\lambda_{1}^{1}\right)   & -\left(  \lambda_{2}^{2}%
+i\lambda_{2}^{1}\right)  \\
\lambda_{1}^{2}-i\lambda_{1}^{1} & \lambda_{2}^{2}-i\lambda_{2}^{1} &
\lambda_{1}^{0}+i\lambda_{1}^{3} & \lambda_{2}^{0}+i\lambda_{2}^{3}\\
-d & \left(  \frac{-i}{2d}\right)  l & 0 & \left(  \frac{-i}{2d}\right)
\left(  m-in\right)  \\
\left(  \frac{-i}{2d}\right)  l & d & \left(  \frac{-i}{2d}\right)  \left(
m-in\right)   & 0\\
0 & \left(  \frac{-i}{2d}\right)  \left(  m+in\right)   & -d & -\left(
\frac{-i}{2d}\right)  l\\
\left(  \frac{-i}{2d}\right)  \left(  m+in\right)   & -\left(  \frac{-i}%
{2d}\right)  l & -\left(  \frac{-i}{2d}\right)  l & d
\end{bmatrix}
=%
\begin{bmatrix}
J_{1} & J_{2}\\
B_{11} & B_{21}\\
B_{12} & B_{22}%
\end{bmatrix}
\end{equation}
where%
\begin{subequations}
\begin{align}
J_{1} &  =%
\begin{bmatrix}
\lambda_{1}^{0}-i\lambda_{1}^{3} & \lambda_{2}^{0}-i\lambda_{2}^{3}\\
\lambda_{1}^{2}-i\lambda_{1}^{1} & \lambda_{2}^{2}-i\lambda_{2}^{1}%
\end{bmatrix}
,J_{2}=%
\begin{bmatrix}
-\left(  \lambda_{1}^{2}+i\lambda_{1}^{1}\right)   & -\left(  \lambda_{2}%
^{2}+i\lambda_{2}^{1}\right)  \\
\lambda_{1}^{0}+i\lambda_{1}^{3} & \lambda_{2}^{0}+i\lambda_{2}^{3}%
\end{bmatrix}
,\\
B_{11} &  =%
\begin{bmatrix}
-d & \left(  \frac{-i}{2d}\right)  l\\
\left(  \frac{-i}{2d}\right)  l & d
\end{bmatrix}
,B_{21}=%
\begin{bmatrix}
0 & \left(  \frac{-i}{2d}\right)  \left(  m-in\right)  \\
\left(  \frac{-i}{2d}\right)  \left(  m-in\right)   & 0
\end{bmatrix}
,\\
B_{12} &  =%
\begin{bmatrix}
0 & \left(  \frac{-i}{2d}\right)  \left(  m+in\right)  \\
\left(  \frac{-i}{2d}\right)  \left(  m+in\right)   & -\left(  \frac{-i}%
{2d}\right)  l
\end{bmatrix}
,B_{22}=%
\begin{bmatrix}
-d & -\left(  \frac{-i}{2d}\right)  l\\
-\left(  \frac{-i}{2d}\right)  l & d
\end{bmatrix}
.
\end{align}
We are now ready to check the costable conditions in Eqs.(\ref{a}), (\ref{b})
and (\ref{c}). If the common eigenvector $\nu$ exists for some ADHM data, then
the dimension of $($Ker $\beta$/ Im $\alpha)$ will vary frm point to point on
$CP^{3}$. For these cases, the holomorphic vector bundle description on
$CP^{3}$ will break down. For simplicity, let $z=1$ and Eqs.(\ref{c}) gives
\end{subequations}
\begin{equation}
\det\left(  J_{1}+wJ_{2}\right)  =0
\end{equation}
or%
\begin{align}
&  \left[  \left(  \lambda_{1}^{0}-i\lambda_{1}^{3}\right)  \left(
\lambda_{2}^{2}-i\lambda_{2}^{1}\right)  -\left(  \lambda_{2}^{0}-i\lambda
_{2}^{3}\right)  \left(  \lambda_{1}^{2}-i\lambda_{1}^{1}\right)  \right]
\nonumber\\
&  +\left[
\begin{array}
[c]{c}%
\left(  \lambda_{1}^{0}-i\lambda_{1}^{3}\right)  \left(  \lambda_{2}%
^{0}+i\lambda_{2}^{3}\right)  -\left(  \lambda_{1}^{2}+i\lambda_{1}%
^{1}\right)  \left(  \lambda_{2}^{2}-i\lambda_{2}^{1}\right)  \\
-\left(  \lambda_{2}^{0}-i\lambda_{2}^{3}\right)  \left(  \lambda_{1}%
^{0}+i\lambda_{1}^{3}\right)  +\left(  \lambda_{2}^{2}+i\lambda_{2}%
^{1}\right)  \left(  \lambda_{1}^{2}-i\lambda_{1}^{1}\right)
\end{array}
\right]  w\nonumber\\
&  +\left[  -\left(  \lambda_{1}^{2}+i\lambda_{1}^{1}\right)  \left(
\lambda_{2}^{0}+i\lambda_{2}^{3}\right)  +\left(  \lambda_{2}^{2}+i\lambda
_{2}^{1}\right)  \left(  \lambda_{1}^{0}+i\lambda_{1}^{3}\right)  \right]
w^{2}\nonumber\\
&  =0,
\end{align}
which can be written as%

\begin{equation}
\left(  n+im\right)  w^{2}+\left(  2il\right)  w+\left(  n-im\right)  =0
\end{equation}
whose solutions are%
\begin{equation}
w=\frac{1}{n+im}\left[  -il\pm\sqrt{-\left(  l^{2}+m^{2}+n^{2}\right)
}\right]  .
\end{equation}
We now examine Eq.(\ref{a}) which can be written as%
\begin{equation}%
\begin{bmatrix}
-d+x & \left(  \frac{-i}{2d}\right)  \left[  l+w\left(  m-in\right)  \right]
\\
\left(  \frac{-i}{2d}\right)  \left[  l+w\left(  m-in\right)  \right]  & d+x
\end{bmatrix}
v=0.
\end{equation}
The existence of eigenvector implies%
\begin{equation}
x^{2}-d^{2}+\frac{1}{4d^{2}}\left[  l+w\left(  m-in\right)  \right]  ^{2}=0.
\end{equation}
The condition of Eq.(\ref{b}) can be written as
\begin{equation}%
\begin{bmatrix}
y-wd & \left(  \frac{-i}{2d}\right)  \left(  m+in-wl\right) \\
\left(  \frac{-i}{2d}\right)  \left(  m+in-wl\right)  & y+wd
\end{bmatrix}
v=0.
\end{equation}
The existence of eigenvector implies%
\begin{equation}
y^{2}-w^{2}d^{2}+\frac{1}{4d^{2}}\left[  m+in-wl\right]  ^{2}=0.
\end{equation}
The solutions for $x$ and $y$ are%
\begin{align}
x  &  =\pm\sqrt{d^{2}-\frac{1}{4d^{2}}\left[  l+w\left(  m-in\right)  \right]
^{2}},\\
y  &  =\pm\sqrt{w^{2}d^{2}-\frac{1}{4d^{2}}\left(  m+in-wl\right)  ^{2}}.
\end{align}
Finally we have to identify the three different forms of the common
eigenvector%
\begin{equation}%
\begin{bmatrix}
\left(  \frac{-i}{2d}\right)  \left[  l+w\left(  m-in\right)  \right] \\
d-x
\end{bmatrix}
\sim%
\begin{bmatrix}
\left(  \frac{-i}{2d}\right)  \left(  m+in-wl\right) \\
wd-y
\end{bmatrix}
\sim%
\begin{bmatrix}
\lambda_{2}^{0}-i\lambda_{2}^{3}-w\left(  \lambda_{2}^{2}+i\lambda_{2}%
^{1}\right) \\
-\left(  \lambda_{1}^{0}-i\lambda_{1}^{3}\right)  +w\left(  \lambda_{1}%
^{2}+i\lambda_{1}^{1}\right)
\end{bmatrix}
.
\end{equation}

\subsubsection{Example One}

For the first sample solution, we choose the moduli $\lambda_{1}^{1}%
=\lambda_{1}^{2}=\lambda_{1}^{3}=\lambda_{2}^{0}=\lambda_{2}^{2}=\lambda
_{2}^{3}=0$, then we get $l=0,$ $n=0$ and $m=\lambda_{1}^{0}\lambda_{2}^{1}$.
With these inputs, $w=\frac{1}{im}\left[  0\pm\sqrt{-m^{2}}\right]  =\pm1$ and
the constraints from common eigenvector become%
\begin{equation}%
\begin{bmatrix}
\pm\left(  \frac{-i}{2d}\right)  m\\
d-\sqrt{d^{2}-\frac{m^{2}}{4d^{2}}}%
\end{bmatrix}
\sim%
\begin{bmatrix}
\left(  \frac{-i}{2d}\right)  m\\
\pm d-\sqrt{d^{2}-\frac{m^{2}}{4d^{2}}}%
\end{bmatrix}
\sim%
\begin{bmatrix}
\mp i\lambda_{2}^{1}\\
-\lambda_{1}^{0}%
\end{bmatrix}
.
\end{equation}
If we choose $d^{2}-\frac{m^{2}}{4d^{2}}=0$, we have%
\begin{equation}
m=2d^{2}=\lambda_{1}^{0}\lambda_{2}^{1}\text{, \ }\lambda_{1}^{0}=-\lambda
_{2}^{1}\text{, }x=y=0. \label{mod}%
\end{equation}
Let's set $\lambda_{1}^{0}=a,\lambda_{2}^{1}=-a$ where $a$ is a complex number
and $a\neq0$, then the corresponding solutions of moduli parameters are%
\begin{align}
&
\begin{bmatrix}
\lambda_{1} & \lambda_{2}\\
y_{11} & y_{12}\\
y_{12} & y_{22}%
\end{bmatrix}
=%
\begin{bmatrix}
\lambda_{1}^{0}-i\lambda_{1}^{3} & -\left(  \lambda_{1}^{2}+i\lambda_{1}%
^{1}\right)  & \lambda_{2}^{0}-i\lambda_{2}^{3} & -\left(  \lambda_{2}%
^{2}+i\lambda_{2}^{1}\right) \\
\lambda_{1}^{2}-i\lambda_{1}^{1} & \lambda_{1}^{0}+i\lambda_{1}^{3} &
\lambda_{2}^{2}-i\lambda_{2}^{1} & \lambda_{2}^{0}+i\lambda_{2}^{3}\\
-d & 0 & \left(  \frac{-i}{2d}\right)  l & \left(  \frac{-i}{2d}\right)
\left(  m-in\right) \\
0 & -d & \left(  \frac{-i}{2d}\right)  \left(  m+in\right)  & -\left(
\frac{-i}{2d}\right)  l\\
\left(  \frac{-i}{2d}\right)  l & \left(  \frac{-i}{2d}\right)  \left(
m-in\right)  & d & 0\\
\left(  \frac{-i}{2d}\right)  \left(  m+in\right)  & -\left(  \frac{-i}%
{2d}\right)  l & 0 & d
\end{bmatrix}
\nonumber\\
&  =%
\begin{bmatrix}
a & 0 & 0 & ia\\
0 & a & ia & 0\\
\frac{-i}{\sqrt{2}}a & 0 & 0 & \frac{a}{\sqrt{2}}\\
0 & \frac{-i}{\sqrt{2}}a & \frac{a}{\sqrt{2}} & 0\\
0 & \frac{a}{\sqrt{2}} & \frac{i}{\sqrt{2}}a & 0\\
\frac{a}{\sqrt{2}} & 0 & 0 & \frac{i}{\sqrt{2}}a
\end{bmatrix}
,a\neq0. \label{mo}%
\end{align}
Note that since $\lambda_{2}^{1}\neq0$, this set of ADHM data is outside of
CFTW case considered in section IV.A which was shown to be a locally free
case. We thus have discovered that, for points $[x:y:z:w]=[0:0:1:\pm1]$ on
$CP^{3}$ and the ADHM data given in Eq.(\ref{mo}), the vector bundle
description of $SL(2,C)$ $2$-instanton on $CP^{3}$ breaks down, and one is led
to use sheaf description for these non-compact YM instantons or "instanton
sheaves" on $CP^{3}$\cite{math2}. Note in addition that for the case of
$SU(2)$ $2$-instanton, $\lambda_{1}^{0}$ , $\lambda_{2}^{1}$ and $d$ are real
numbers inconsistent with Eq.(\ref{mod}). So the complete $SU(2)$
$2$-instanton solutions are locally free. This is consistent with the known
vector bundle description of $SU(2)$ $2$-instanton on $CP^{3}$.

In the language of monad construction discussed in section III.B, the map
$\alpha$ fails to be injective for points $[x:y:z:w]=[0:0:1:\pm1]$ on $CP^{3}$
with ADHM data given in Eq.(\ref{mo}). Thus one is led to use sheaf
description for these "instanton sheaves" on $CP^{3}$\cite{math2}.

\subsubsection{Example Two}

\bigskip For the second sample solution, we choose the moduli $\lambda_{1}%
^{1}=\lambda_{1}^{2}=\lambda_{1}^{3}=\lambda_{2}^{0}=\lambda_{2}^{1}%
=\lambda_{2}^{3}=0$, then we get $l=0,$ $n=\lambda_{1}^{0}\lambda_{2}^{2}$ and
$m=0$. With these inputs, $w=\pm i$ and the constraints from common
eigenvector become%
\[%
\begin{bmatrix}
\pm\left(  \frac{-i}{2d}\right)  in\\
id-\sqrt{-d^{2}+\frac{n^{2}}{4d^{2}}}%
\end{bmatrix}
\sim%
\begin{bmatrix}
\left(  \frac{-i}{2d}\right)  n\\
\pm d-\sqrt{d^{2}-\frac{n^{2}}{4d^{2}}}%
\end{bmatrix}
\sim%
\begin{bmatrix}
\mp i\lambda_{2}^{2}\\
-\lambda_{1}^{0}%
\end{bmatrix}
.
\]
If we choose $d^{2}-\frac{n^{2}}{4d^{2}}=0$, we have%
\begin{equation}
n=2d^{2}=\lambda_{1}^{0}\lambda_{2}^{2}\text{ , \ }\lambda_{1}^{0}%
=-\lambda_{2}^{2}\text{, \ }x=y=0.
\end{equation}
Let's set $\lambda_{1}^{0}=a,\lambda_{2}^{2}=-a$ where $a$ is a complex number
and $a\neq0$, then the corresponding solutions of moduli parameters are%
\begin{align}
&
\begin{bmatrix}
\lambda_{1} & \lambda_{2}\\
y_{11} & y_{12}\\
y_{12} & y_{22}%
\end{bmatrix}
\symbol{126}%
\begin{bmatrix}
\lambda_{1}^{0}-i\lambda_{1}^{3} & -\left(  \lambda_{1}^{2}+i\lambda_{1}%
^{1}\right)  & \lambda_{2}^{0}-i\lambda_{2}^{3} & -\left(  \lambda_{2}%
^{2}+i\lambda_{2}^{1}\right) \\
\lambda_{1}^{2}-i\lambda_{1}^{1} & \lambda_{1}^{0}+i\lambda_{1}^{3} &
\lambda_{2}^{2}-i\lambda_{2}^{1} & \lambda_{2}^{0}+i\lambda_{2}^{3}\\
-d & 0 & \left(  \frac{-i}{2d}\right)  l & \left(  \frac{-i}{2d}\right)
\left(  m-in\right) \\
0 & -d & \left(  \frac{-i}{2d}\right)  \left(  m+in\right)  & -\left(
\frac{-i}{2d}\right)  l\\
\left(  \frac{-i}{2d}\right)  l & \left(  \frac{-i}{2d}\right)  \left(
m-in\right)  & d & 0\\
\left(  \frac{-i}{2d}\right)  \left(  m+in\right)  & -\left(  \frac{-i}%
{2d}\right)  l & 0 & d
\end{bmatrix}
\nonumber\\
&  =%
\begin{bmatrix}
a & 0 & 0 & a\\
0 & a & -a & 0\\
\frac{-i}{\sqrt{2}}a & 0 & 0 & \frac{-ia}{\sqrt{2}}\\
0 & \frac{-i}{\sqrt{2}}a & \frac{ia}{\sqrt{2}} & 0\\
0 & \frac{-ia}{\sqrt{2}} & \frac{i}{\sqrt{2}}a & 0\\
\frac{ia}{\sqrt{2}} & 0 & 0 & \frac{i}{\sqrt{2}}a
\end{bmatrix}
,a\neq0. \label{mo2}%
\end{align}
Note that since $\lambda_{2}^{2}\neq0$, this set of ADHM data is again outside
of CFTW case considered in section IV.A. Eq.(\ref{mo2}) gives the second
example of sheaf description of the$SL(2,C)$ $2$-instanton solution on
$CP^{3}$. Note again that for the case of $SU(2)$ $2$-instanton, $\lambda
_{1}^{0}$ , $\lambda_{2}^{2}$ and $d$ are real numbers inconsistent with
Eq.(\ref{mod}). We thus have discovered that, for points $[x:y:z:w]=[0:0:1:\pm
i]$ on $CP^{3}$ and the ADHM data given in Eq.(\ref{mo2}), the vector bundle
description of $SL(2,C)$ $2$-instanton on $CP^{3}$ breaks down. Note that for
this case, the map $\alpha$ in the monad construction fails to be injective
for points $[x:y:z:w]=[0:0:1:\pm i]$ on $CP^{3}$ with ADHM data given in
Eq.(\ref{mo2}).

To further catch the geometric picture we remark that outside the above points
$[x:y:z:w]=[0:0:1:\pm1]$ or $[0:0:1:\pm i]$ (or setting $w=1$ instead of $z=1$
with similar formulas) the vector bundle description is valid, and more
generally for any other biquaternion ADHM data a certain set of finitely many
points can be found similarly as above, so that the corresponding vector
bundle description is necessarily valid outside these finitely many points
(this does not mean, however, that the vector bundle description has to be
broken at these finitely many points). On the other hand, as we have shown
that a CFTW $k$-instanton solution gives a global vector bundle on $CP^{3}$,
its small perturbation as a general $SL(2,C)$ $2$-instanton solution shall
remain a global vector bundle solution (which is due to the fact that if a
system of linear equations have no common solution with a certain parameter
(i.e. a vector bundle case), then the same can be said with all nearby
parameters corresponding to the perturbations; the parameters here can be
specified by our ADHM data).

We conclude that for a certain proper subset of our biquaternion ADHM data
(including the examples above), the vector bundle description breaks down only
at those finitely many points (whose positions depend on the details of the
ADHM data). However for ADHM data outside this proper subset the vector bundle
description remains valid on the whole $CP^{3}$. Mathematically \cite{math2}
the breakdown of the vector bundle description is related to the third Chern
number $c_{3}$ of the obtained sheaf, which could be nonzero in the sheaf case
in contrast to the vector bunlde case in which $c_{3}$ is necessarily zero
because the bundle is of rank two (two dimensional).

Secondly, as remarked in Introduction ADHM construction highly depends on one
to one correspondence between ASD connections on the one side and certain
holomorphic objects on the other side-twistor space, which is mainly
accomplished by Penrose-Ward transform. By using the knowledge and information
on the twistor side one may therefore reach an understanding for ASD
connections. This idea works most effectively for the vector bundle case to
which the $SU(2)$ instantons perfectly belong. However in the present
$SL(2,C)$ case the holomorphic objects are no longer merely vector bundles on
the twistor space as we have discussed in this paper, which renders the
corresponding transformation in between less clarified. For instance, a vector
bundle on $CP^{3}$ can descend down to $S^{4}$ if and only if its set of
jumping lines does not include any fiber of the fibration map $CP^{3}%
\rightarrow S^{4}$. It is a worthy work to examine the singularities of the
Penrose-Ward transformed object on $S^{4}$. Presumably the singularities may
appear when the preceding jumping-line condition is not met, or when the
holomorphic object on $CP^{3}$ is a sheaf instead of a vector bundle. Since
the cases vary and the nature of the problem appears quite different from case
to case, we shall leave the study of singularites on $S^{4}$ to a future work.

Finally it is a natural question to ask whether our biquaternion ADHM
solutions give all solutions to complex ADHM equations. As remarked in the
Introduction the number $16k-6$ of parameters obtained by our biquaternion
ADHM method is basically the expected number in mathematics. Yet an interested
reader will soon find the solution for $k=1$ with $I_{1}=[1\,\,0]$,
$I_{2}=[0\,\,1]$, $J_{1}=J_{2}=0$ and all $B_{lm}=0$ \cite{math2} seem to lie
outside the biquaternion ADHM data. Now if one takes $I_{1}=[t\,\,0]$,
$I_{2}=[0\,\,t]$, $J_{1}=-I_{1}^{\dagger}$, $J_{2}=I_{2}^{\dagger}$ and all
$B_{lm}=0$, which is seen to be a biquaternion ADHM solution and is equivalent
to $(B_{lm},gI_{m},J_{m}{g^{-1}})$ in general for any nonzero complex number
$g$, then one sees, by setting $g=t^{-1}$ and letting $t\rightarrow0$ the
above biquaternion solution under equivalence indeed reproduces the above
(non-biquaternion) solution. It is to be expected that by a certain limiting
procedure the biquaternion ADHM solutions (up to equivalence) reproduce all
solutions to complex ADHM equations.

\section{Conclusion}

In the ADHM construction of $SU(2)$ YM instantons, one establishs an one to
one correspondence between anti-self-dual $SU(2)$ connections on $S^{4}$ and
global holomorphic vector bundles of rank two on $CP^{3}$ satisfying certain
reality conditions. In this paper, we try to extend this correspondence to the
case of non-compact $SL(2,C)$ YM instanton. As the first step of this program,
the $SL(2,C)$ YM instanton solutions constructed recently by the biquaternion
method were shown to satisfy the complex version of the ADHM equations and the
monad construction. We can then identify the complex ADHM data for these
$SL(2,C)$ YM instanton along the calculation.

The next step was to calculate the costable and stable conditions of these
ADHM data. For the case of $SL(2,C)$ CFTW $k$-instanton solutions with $10k$
moduli parameters, although there exist twistor lines which are jumping lines
on $S^{4}$, the corresponding ADHM data are locally free and the vector bundle
description of $SL(2,C)$ CFTW $k$-instanton on $CP^{3}$ remains valid as in
the case of $SU(2)$ instantons. We then proceed to calculate the second case
of complete known $SL(2,C)$ $2$-instanton solutions with $26$ moduli
parameters. We discover that, for some points on $CP^{3}$ and some subset of
the complex ADHM data of $SL(2,C)$ $2$-instanton solutions, the vector bundle
description of $SL(2,C)$ $2$-instanton on $CP^{3}$ breaks down, and one is led
to use sheaf description for these non-compact YM instantons or "instanton
sheaves" on $CP^{3}$.

Although the existence of instanton sheaf has been discussed previously, their
explicit constructions have not been worked out yet. We expect that the
explicit forms of\bigskip\ the $SL(2,C)$ YM instanton sheaf solutions
constructed in this paper will be helpful to the further developments on this
subject both physically and mathematically.

\begin{acknowledgments}
The work of J.C. Lee is supported in part by the Ministry of Science and
Technology and S.T. Yau center of NCTU, Taiwan. The work of I-H. Tsai has been
possible due to an opportunity for him to visit S.T. Yau center of NCTU to
which he owes his thanks.
\end{acknowledgments}


\begin{thebibliography}{99}                                                                                               %
\bibitem {U(1)}G. 't Hooft, "Computation of the quantum effects due to a
four-dimensional pseudoparticle", Phys. Rev. D 14 (1976) 3432. G. 't Hooft,
"Symmetry breaking through Bell-Jackiw anomalies", Phys. Rev. Lett. 37 (1976) 8.

\bibitem {the}C. Callan Jr., R. Dashen, D. Gross, "The structure of the gauge
theory vacuum", Phys. Lett. B 63 (1976) 334; "Toward a theory of the strong
interactions", Phys. Rev. D 17 (1978) 2717. R. Jackiw, C. Rebbi, "Vacuum
periodicity in a Yang-Mills quantum theory", Phys. Rev. Lett. 37 (1976) 172.

\bibitem {5}S.K. Donaldson and P.B. Kronheimer, \textquotedblright The
Geometry of Four Manifolds\textquotedblright, Oxford University Press (1990).

\bibitem {BPST}A. Belavin, A. Polyakov, A. Schwartz, Y. Tyupkin,
"Pseudo-particle solutions of the Yang-Mills equations", Phys. Lett. B 59
(1975) 85.

\bibitem {CFTW}{E.F. Corrigan, D.B. Fairlie, Phys. Lett. 67B (1977)69; G.
'tHooft, Phys. Rev. Lett., 37 (1976) 8; F. Wilczek, in "Quark Confinement and
Field Theory", Ed. D.Stump and D. Weingarten, John Wiley and Sons, New York
(1977).}

\bibitem {JR}R. Jackiw, C. Rebbi, "Conformal properties of a Yang-Mills
pseudoparticle", Phys. Rev. D 14 (1976) 517; R. Jackiw, C. Nohl and C. Rebbi,
"Conformal properties of pseu-doparticle con gurations", Phys. Rev. D 15
(1977) 1642.

\bibitem {ADHM}M. Atiyah, V. Drinfeld, N. Hitchin, Yu. Manin, "Construction of
instantons", Phys. Lett. A 65 (1978) 185.

\bibitem {CSW}N. H. Christ, E. J. Weinberg and N. K. Stanton, "General
Self-Dual Yang-Mills Solutions", Phys. Rev. D 18 (1978) 2013. V. Korepin and
S. Shatashvili, "Rational parametrization of the three instanton solutions of
the Yang-Mills equations", Math. USSR Izversiya 24 (1985) 307.

\bibitem {JR2}R. Jackwi and C. Rebbi, Phys. Lett. 67B (1977) 189. C. W.
Bernard, , N. H. Christ, A. H. Guth and E. J. Weinberg, Phys. Rev. D16 (1977) 2967.

\bibitem {Ann}S. H. Lai, J. C. Lee and I. H. Tsai, "Biquaternions and ADHM
Construction of Non-Compact SL(2,C) Yang-Mills Instantons", Annals Phys. 361
(2015) 14.

\bibitem {math2}I. Frenkel and M. Jardim, "Complex ADHM equations and sheaves
on $P^{3}$", Journal of Algebra 319 (2008) 2913-2937. J. Madore, J.L. Richard
and R. Stora, "An Introduction to the Twistor Programme", Phys. Rept. 49, No.
2 (1979) 113-130.

\bibitem {math3}M. Jardim and M. Verbitsky, "Trihyperkahler reduction and
instanton bundles on $CP^{3}$", Compositio Math. 150 (2014) 1836.

\bibitem {Lee}K. L. Chang and J. C. Lee, "On solutions of self-dual SL(2,C)
gauge theory", Chinese Journal of Phys. Vol. 44, No.4 (1984) 59. J.C. Lee and
K. L. Chang, "SL(2,C) Yang-Mills Instantons", Proc. Natl. Sci. Counc. ROC (A),
Vol 9, No 4 (1985) 296.

\bibitem {Donald}S. Donaldson, "Instantons and Geometric Invariant Theory",
Comm. Math. Phys. 93 (1984) 453--460.

\bibitem {WY}{Tai Tsun Wu and Chen Ning Yang, Phys. Rev. D12, 3843 (1975);
Phys. Rev.D13, (1976) 3233.}

\bibitem {Ham}W. R. Hamilton, \textquotedblright Lectures on
Quaternions\textquotedblright, Macmillan \& Co, Cornell University Library (1853).

\bibitem {LLT4}S. H. Lai, J. C. Lee and I. H. Tsai, "Sheaf lines of Yang-Mills
Instanton Sheaves", arXiv: 1708.02853.
\end{thebibliography}
\end{document}